# Dependence of the Critical Adsorption Point on Surface and Sequence Disorders for Self-Avoiding Walks Interacting with a Planar Surface

Jesse D. Ziebarth[1,] Yongmei Wang[1*], Alexey Polotsky[2], Mengbo Luo[3]

[1]Department of Chemistry, the University of Memphis, Memphis, Tennessee 38152, USA
[2] Sérvice de Physique de l'Etat Condensé CEA Saclay, 91191 Gif-sur-Yvette Cedex, France.
[3]Department of Physics, Zhejiang University, Hangzhou 310027, P. R. China

**Abstract**. The critical adsorption point (CAP) of self-avoiding walks (SAW) interacting with a planar surface with surface disorder or sequence disorder has been studied. We present theoretical equations, based on ones previously developed by Soteros and Whittington (*J. Phys. A.: Math. Gen.* **2004**, *37*, R279-R325), that describe the dependence of CAP on the disorders along with Monte Carlo simulation data that are in agreement with the equations. We also show simulation results that deviate from the equations when the approximations used in the theory break down. Such knowledge is the first step toward understanding the correlation of surface disorder and sequence disorder during polymer adsorption.

*Corresponding author: email: ywang@memphis.edu. Tel: 901-678-2629. Fax: 901-678-3447.

# 1. Introduction

Adsorption of polymers on surfaces plays a key role in many technological applications and is also relevant to many biological processes. As a result, it has been studied for more than three decades[1] and continues to receive intense interest.[2] The field is rich and contains a wide variety of topics, from equilibrium properties of adsorbed layers and conformations of adsorbed polymer chains to dynamic properties and non-equilibrium processes in adsorption.[2] For polymer adsorption on planar surfaces, it is well-known that there exists a critical adsorption point (CAP) that marks the transition of a polymer chain, in contact with a surface, from a non-adsorbed state to an adsorbed state.[3] Scaling laws for a variety of quantities below, above and at the CAP for a homopolymer in contact with a planar surface were developed by Eisenriegler, Kremer, and Binder (EKB).[4] For example, when the chain goes from a non-adsorbed state to an adsorbed state, the energy of the chain $E$ changes from an intensive variable independent of chain length $N$ to an extensive variable dependent on $N$. At the CAP, $E$ is expected to scale with $N^\phi$ where $\phi$ is the crossover exponent. Numerical studies, including exact enumeration,[5] the scanning method[6,7] and the multiple Markov chain method[8] have been performed to determine the location of the CAP and the crossover exponent $\phi$. The values reported are however not completely in agreement with each other and are still under debate, especially the crossover exponent $\phi$. The disagreement may be traced, as suggested by a recent article,[9] to different methods used for determining the CAP and the crossover exponent $\phi$.

While many studies focused on adsorption of homopolymers on planar homogeneous surfaces, adsorption of polymers on chemically or physically heterogeneous surfaces has also received a fair amount of studies.[10-20] Some were inspired by specific applications such as segregation of polymer chains on patterned surfaces,[10] or pattern transfer via surface adsorption,[21,22] others were motivated by a desire to understand how the presence of surface or sequence disorders may influence adsorption.[13,14,16,17,23-25] For example, Sebastian and Sumithra developed an analytical theory of the adsorption of Gaussian chains on random surfaces using Gaussian variational approach.[24,25] They took surface heterogeneity into account by modifying de Genne's adsorption boundary condition and analyzed influence of randomness on the conformation of the adsorbed chains. Adsorption of heteropolymers on heterogeneous surfaces, in particular, has been studied because of its relevance to molecular recognition in biological process. The concept of "pattern matching" was proposed[26] and has been investigated with different approaches.[12,20,26,27] Muthukumar for example derived an equation for the critical condition of adsorption of a polyelectrolyte to an oppositely charged patterned surface.[26] Golumbfski et al.[12] showed that a statistical blocky chain was selectively adsorbed on a patchy surface while a statistically alternating chain was selectively adsorbed on an alternating surface. Jayaraman et al.[19] described a simulation method to design surfaces for recognizing specific monomer sequences in heteropolymers. Recently Polotsky et al[18] considered adsorption of Gaussian heteropolymer chains onto heterogeneous surface. They found that the presence of correlations between sequence and surface heterogeneity always enhances adsorption. However, the dependence of the critical adsorption point on either surface disorder or sequence disorder is not well-understood. Lack of this knowledge hampers further understanding on the correlation between sequence disorder and surface disorder during adsorption.

Here we present theoretical equations that describe the dependence of CAP on the surface disorder or sequence disorder, along with Monte Carlo simulation data in agreement with the derived equations. The current study does not address the correlation between sequence disorder and surface disorder. We only consider cases where the disorder is either present randomly on the surface (i.e. adsorption of homopolymers on random heterogeneous surface) or on the sequence (i.e., adsorption of random copolymer on homogeneous surface). The correlation between sequence disorder and surface disorder will be the subject of future publications. In the following, we first present the theory that predicts the dependence of CAP on surface disorder and sequence disorder. Then we present details of Monte Carlo simulation methods used to determine the CAP, followed by simulation data that agree with the derived equations. Finally, we discuss implications of these results on practical applications such as chromatographic separations of polymers.

## 2. Theory

### 2.1 Adsorption of a homopolymer on a homogeneous surface

We first consider adsorption of a homopolymer chain on a homogeneous surface. This can be represented by a self-avoiding walk (SAW) in a three-dimensional lattice interacting with a plane and restricted to lie on one side of the plane. The vertices of the walks interact with the surface sites with an attractive energy $\varepsilon_w$. The partition function for a $N$-step SAW interacting with a homogeneous surface is given by

$$Z_{\text{homo}}(N, \varepsilon_w) = \sum_v c_N(v) \exp(\varepsilon_w v) \tag{1}$$

where $c_N(v)$ is the number of SAWs that lie above the surface with $v$ visits to the surface. Hammersley *et al.*[28] have shown that the model exhibits a phase transition at a critical adsorption energy, $\varepsilon_c$, with a desorbed state for $\varepsilon_w < \varepsilon_c$, and an adsorbed state for $\varepsilon_w > \varepsilon_c$. They have shown that the limiting monomer free energy $f(\varepsilon_w)$

$$f(\varepsilon_w) = \lim_{N \to \infty} \frac{1}{N} \log Z_{\text{homo}}(N, \varepsilon_w) \tag{2}$$

exists and is a convex non-decreasing continuous function of $\varepsilon_w$. Moreover, $f(\varepsilon_w)=\kappa$ for $\varepsilon_w \leq 0$, where $\kappa$ is the lattice connective constant, and $f(\varepsilon_w)$ is a strictly increasing function of $\varepsilon_w$ when $\varepsilon_w > \varepsilon_c$. Therefore, $f(\varepsilon_w)$ is non-analytic at $\varepsilon_w = \varepsilon_c$. $\varepsilon_c$ has also been determined to be greater than zero and, based on the best-known connective constant for the simple cubic lattice[29], to have an upper bound of 0.5738. The lattice connective constant $\kappa$ is also the limiting monomer free energy of the SAWs in bulk solution. Hence the CAP can be understood as the condition where the limiting monomer free energy of a chain attached to the surface becomes equal to the limiting monomer free energy of the chain in the bulk solution.

### 2.2 Adsorption of a homopolymer on a random heterogeneous surface

Now we consider the adsorption of a homopolymer interact with a heterogeneous surface consisting of two types of surface sites, $A$ and $B$. The interaction energy of the vertices with the two surface sites are $\varepsilon_w^A$ and $\varepsilon_w^B$. Following Soteros and Whittington[23], and express the partition function of a $N$-step SAW interacting with a heterogeneous surface that consists of A and B surface sites as:

$$Z_{het}(f_A, f_B) = \sum_v \sum_{v(A)=0}^{v} c_N(v) \binom{v}{v(A)} f_A^{v(A)} \exp(\varepsilon_w^A v(A)) f_B^{v(B)} \exp(\varepsilon_w^B v(B)) \qquad (3)$$

where $c_N(v)$ is the number of walks that have $v$ surface contacts, $v(A)$ is the number of monomers interacting with the A sites, and $v(B)$ is the number of monomers interacting with the B sites, $f_A$ and $f_B = 1 - f_A$ are the fractions of A and B sites on the surface, respectively. Here the partition function is averaged over random distributions of the surface sites, i.e. the so called annealed approximation. Physically the annealed disorder means that the type of surface sites may change while the system attains equilibrium state. However, it has been previously suggested[11,15] that the annealed approximation is valid if the chain can visit a large area of the surface and hence samples all distributions of surface patterns. Furthermore, the surface sites are randomly distributed. If there is a correlation between surface disorders, such as those present in patchy surface or alternating surface, then Eq. (3) will not be valid, as Eq. (3) gives equal weight to all possible surface labelings, while correlations restrict possible labelings. Summing over $v(A)$, equation (3) can be simplified to

$$Z_{het}(f_A, f_B) = \sum_v c_N(v)\left(f_A \exp(\varepsilon_w^A) + f_B \exp(\varepsilon_w^B)\right)^v \qquad (4)$$

A comparison of equations (1) and (4) reveals that the partition functions for homogeneous and annealed random heterogeneous surface become equivalent if

$$\exp(\varepsilon_w) = f_A \exp(\varepsilon_w^A) + f_B \exp(\varepsilon_w^B) \qquad (5)$$

From Eq. (5), we derive the following equation that gives the dependence of CAP on the surface disorder:

$$\exp(\varepsilon_w^h(cc)) = (1 - f_B)\exp(\varepsilon_w^A) + f_B \exp(\varepsilon_w^B(cc)) \qquad (6)$$

where $\varepsilon_w^h$(cc) is the CAP of a homopolymer above a homogeneous surface, $\varepsilon_w^B$(cc) is the CAP of a homopolymer above a heterogeneous surface while the surface interaction energy $\varepsilon_w^A$ held constant. It can be easily seen from this equation, that the dependence of the CAP on the percentage of attractive sites on the surface is not expected to be linear, in contrast to the conclusion drawn by an earlier study.[13] Equation (6) is expected to be valid as long as the two conditions are met: (i) the chain has enough mobility to visit a large area of surface so that the annealed approximation is valid, and (ii) the surface sites are randomly distributed (i.e. uncorrelated).

*2.3 Adsorption of a random heteropolymer on a homogeneous surface*

The same approach can be extended to consider the adsorption of a random heteropolymer interacting with a homogeneous surface. We will use the same notation as in previous section except now $f_A$ and $f_B$ represent fractions of A and B monomers present on the heteropolymer. We will only consider random copolymers composed by A and B monomers. The sequence of a random copolymer can be represented by $\chi = \{\chi_1, \chi_2, \ldots \chi_N\}$ where $\chi_i$ are independently and identically distributed random variables with $\chi_i = A$ with a probability of $f_A$ and $\chi_i = B$ with a probability of $1-f_A$. A sequence order parameter $\lambda$ can be defined to characterize the sequence randomness.[12,27]

$$\lambda = 1 - p_{AB} - p_{BA} \tag{7}$$

where $p_{ij}$ is the nearest neighbor transition probabilities which is the probability that a monomer of type $i$ is followed by a monomer of type $j$. When $\lambda=0$, the sequence is random. When $\lambda>0$, then the sequence is statistically blocky, and when $\lambda<0$, the sequence is statistically alternating. We note that a given random sequence designated by $\chi$ may have non-zero values of $\lambda$. More discussions will be given in the later section.

The partition function of $N$-step SAWs with the given sequence above a homogenous surface is written as:

$$Z_{hetpoly}(f_A, f_B, \chi) = \sum_v C_N(v_A, v_B \mid \chi) \exp(v_A \varepsilon_w^A + v_B \varepsilon_w^B) \tag{8}$$

There are two different ways to average over different distributions of random sequences, namely the annealed average and the quenched average. With the annealed average, the partition function in Eq. (8) is first averaged over different distributions of $\chi$. This then leads to a partition function, $Z_{hetpoly}(f_A, f_B)$, which is exactly the same as in Eq. (3). With the annealed approximation, we derive the same equation as given by Eq. (6) for the CAP of a random heteropolymer interacting with a homogeneous surface, provided that $f_A$ and $f_B$ now represent the fractions of A and B monomers on the chain.

In the following, we will present Monte Carlo simulation data that conform to the two equations and also results that do not conform to the equations because of the invalidation of the approximations used in deriving the equations.

## 3. Monte Carlo Simulation Methods

In our simulations, polymer chains are modeled as SAWs with $N$ vertices on a simple cubic lattice of dimensions $250a \times 250a \times 100a$, where $a$ is the lattice spacing. Each vertex represents a monomer on the polymer chain. Chain lengths studied are in the range of $N = 25$ to 250. There is an impenetrable wall in the $z = a$ plane representing the surface. One monomer, picked randomly from the chain, is first placed on a site adjacent to the wall (in the $z = 2a$ plane). The rest of the chain is then grown using the biased chain insertion method.[30] Monomers that are in the $z = 2a$ plane are considered to be adsorbed on the surface. For all adsorbed monomers, an attractive polymer-surface interaction, $\varepsilon_w$, is applied. The standard chemical potential of the chain (since it does not contain translation entropy), $\mu^0$, is calculated from the Rosenbluth-Rosenbluth weighting factor, $W(N)$, which is given by[30]

$$\beta\mu^0 = -\ln\langle W(N)\rangle = -\ln\left\langle\prod_{i=1}^{N} w_i\right\rangle \text{ and } w_i = \frac{\sum_{j=1}^{z}\exp(-\beta E_j)}{z} \qquad (9)$$

where $z$ is the lattice coordination number ($z = 6$ for simple cubic lattice), $E_j$ is the energy of $i$th inserted monomer in the $j$th potential direction. We note that $\mu^0$ calculated is the free energy per chain, and $\mu^0/N$ is free energy per monomer discussed in equation (2). Typically, the chemical potential is determined based on about twenty million copies of trial chain conformations.

We obtained the standard chemical potentials of a chain with at least one monomer attached to the surface, $\mu_{ads}^0$, and compared that against a chain grown in a bulk solution, $\mu_{bulk}^0$. The bulk solution is modeled by a $100a \times 100a \times 100a$ lattice with periodic boundary conditions applied in all three directions. All chemical potentials calculated are reduced by the Boltzmann factor, $\beta=1/k_BT=1$. A coefficient $K$, similar to partition coefficient if the chain was placed in a pore instead of near a surface, is calculated by $K =\exp(-\Delta\mu^0)$, where $\Delta\mu^0 = \mu_{ads}^0 - \mu_{bulk}^0$. The way we determined the CAP is based on the dependence of $K$ on the chain length $N$ and will be presented in the results section.

Heterogeneous surfaces were modeled by making the $z = a$ plane composed of two different types of sites, which have different values for polymer-surface interactions. The designations $\varepsilon_w^A$ and $\varepsilon_w^B$ will be used to distinguish between interaction energies of different site types. Simulations were performed using surfaces with different fractions of A and B sites. Surfaces were created by randomly assigning each site as A or B based on the probabilities, $p_A$ and $p_B$, where $p_A$ and $p_B$ are, respectively, the desired fractions of A and B sites on the surface. Because of size of the surface, this procedure resulted in the real surface composition percentages matching the desired percentages within 0.1%. For a given surface composition, the surface was randomly created once and was subsequently used in all simulations that determine the chemical potential of a chain above that surface. The surfaces displayed quenched randomness, i.e. the surface pattern remained unchanged throughout the simulations. However, the first bead of chain was placed randomly over the surface during the chain insertion, and hence the chemical potential determined has been averaged over different surface randomness.

Therefore, the annealed approximation used in deriving Eq. (6) was met in the simulations. In a few cases, patchy and alternating surfaces were created by simulating a two-dimensional Ising model at appropriate conditions.

Heteropolymers were modelled as SAWs consisting of two types of monomers, A and B with specified fractions $f_A$ and $f_B=1-f_A$. Chains were created by randomly selecting $N*f_B$ different positions along the chain to be B beads, while the remaining beads were assigned as A beads, ensuring that the chain had the exact composition called for by $f_A$ and $f_B$. The sequence order parameter, $\lambda$, in generated random sequences exhibits a Gaussian distribution with zero mean. Examples of distributions are presented in Figure 1. The longer the chain, the narrower the distribution is. For a given chain length $N$, we typically generate 5000 copies of random sequences with specified $f_A$. Each sequence is then used in biased insertion for 5000 or more copies to obtain the Rosenbluth-Rosenbluth weighting factor. Letting W($N$, $\chi$) stands for the Rosenbluth-Rosenbluth weighting factor obtained for a given sequence $\chi$, the chemical potential of a chain can be obtained using two different averages over sequences:

$$\beta\mu^0_{ads}(N) = -\ln\langle W(N,\chi)\rangle \tag{10}$$

$$\beta\mu^0_{ads}(N) = \langle\beta\mu^0_{ads}(N,\chi)\rangle = -\langle\ln W(N,\chi)\rangle \tag{11}$$

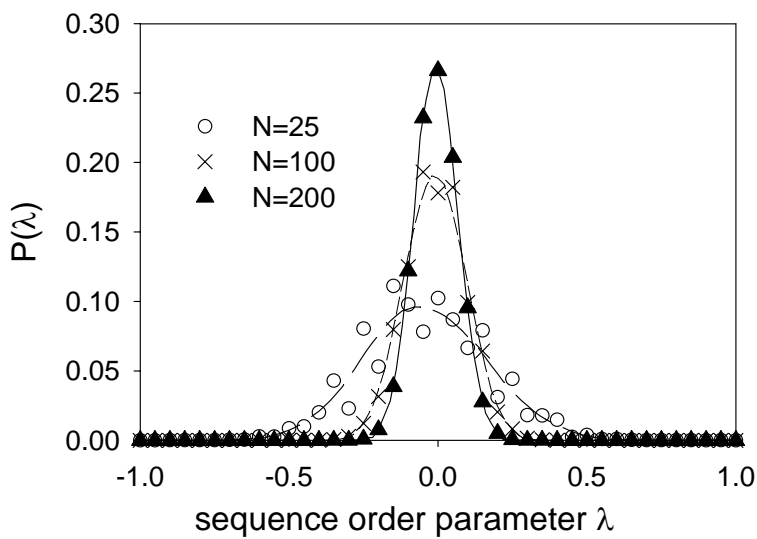

**Figure 1: Distribution of sequence order parameters obtained from 5000 copies of random sequences generated with $f_A = f_B = 0.50$ for three different chain lengths. Lines are smooth fit to the data.**

The first approach is the annealed average, while the second approach is the quenched average. The two chemical potentials calculated differ slightly from each other. More discussion of the quenched versus annealed averages will be given later. For the determination of CAP, we have used annealed chemical potentials.

## 4. Results and Discussion

### 4.1. Method Used to Determine the Critical Adsorption Point

The method we used to determine the CAP follows our earlier papers[31,32] and is briefly sketched out. We obtain the difference in standard chemical potential $\Delta\mu^0$ at different surface interaction $\varepsilon_w$ for a set of chains with different lengths. An example of data is presented in Figure 2(a) for a homopolymer above a homogeneous surface. The lines for different length $N$ nearly intersect at a common point, which is estimated to be at $\varepsilon_c=0.276 \pm 0.005$. A convenient way to identify this intersection point is to plot the standard deviation of all $\Delta\mu^0$, $\sigma(\Delta\mu^0)$, for a given range of chain length studied versus $\varepsilon_w$, which yields a minimum in a plot shown in Figure 2(b). The minimum identified is directly related to the critical condition point employed in liquid chromatography at the critical condition (LCCC) [32-34]. In LCCC, the critical condition was defined as the co-elution point of homopolymers with different molecular weights, which, corresponding to computer simulation, is the point where $K$ has least dependence on chain length. If $K$ is truly independent of chain length, then $\sigma(\Delta\mu^0)$ will be zero and will be the minimum in a plot in Figure 2(b). The critical condition point bracketed in this fashion depends slightly on the range of chain length included in the calculation of $\sigma(\Delta\mu^0)$. However, in the current study we fixed the range of chain lengths used.

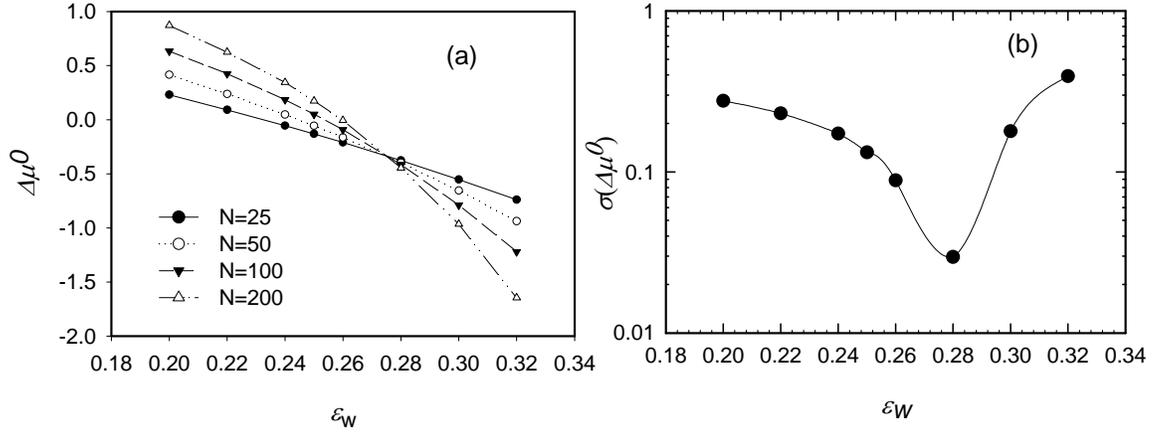

**Figure 2: (a) Plot of $\Delta\mu^0$ versus $\varepsilon_w$ for SAW chains with $N$ =25, 50, 100 and 200 above a homogeneous surface. The critical adsorption point is identified as the common intersection point, $\varepsilon_w(cc)$=0.276±0.005. (b) Plot of deviation in $\Delta\mu^0$ for the given range of $N$ versus $\varepsilon_w$. The minimum in the plot is the critical adsorption point.**

Since this common intersection point does not occur at $\Delta\mu^0$ =0, one may wonder if it is the critical adsorption point discussed in the literature. We have applied the same method for random walks above a planar surface in simple cubic lattice[31]. The intersection point found was at $\varepsilon_c$ = 0.183± 0.002, in excellent agreement with expected CAP for random-walks, $\varepsilon_c$ = -ln(5/6)= 0.1823.[1] On the other hand, CAP could be understood as the point where the limiting monomer free energy for a chain attached to the surface $f(\varepsilon)$ equals to the limiting monomer free energy of an unattached chain in the bulk solution. Therefore, we may define a CAP at a finite chain length, $\varepsilon_c(N)$, at which $\Delta\mu^0(N)=0$. From Figure 2(a), we extract such $\varepsilon_c(N)$. This $\varepsilon_c(N)$ is expected to depend on $N$ in a scaling law, $\varepsilon_c(N) = \varepsilon_c(\infty) - \alpha N^{-\phi}$, and $\varepsilon_c(\infty)$ is the CAP at infinite chain length limit. Assuming $\phi = 0.5$, Figure 3 shows the linear fitting of $\varepsilon_c(N)$ versus $N^{-0.5}$ which yields $\varepsilon_c(\infty) = 0.274 \pm 0.005$. The $\varepsilon_c(\infty)$ identified is within the error bars of the common intersection point.

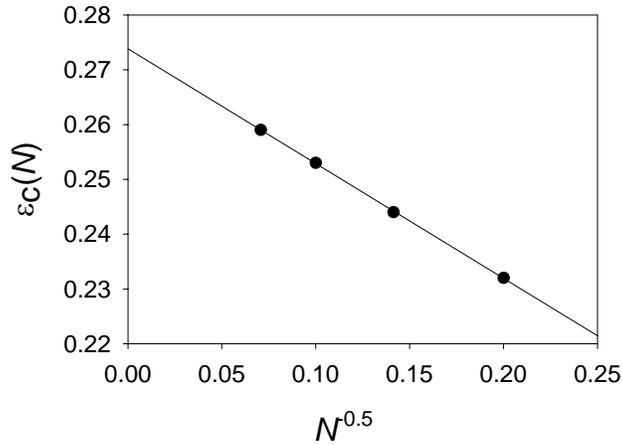

**Figure 3: Plot of $\varepsilon_c(N)$ versus $N^{-0.5}$ where $\varepsilon_c(N)$ is extracted from figure 1(a) as the point when $\Delta\mu(N) = 0$. The extrapolated $\varepsilon_c(\infty) = 0.274 \pm 0.005$.**

The CAP of SAWs in simple cubic lattice has been studied by others.[6-8] The reported literature value for the CAP of SAWs on the simple cubic lattice ranged from ~0.37 by Ma et al.[35] down to $0.288 \pm 0.02$ by Janse van Rensburg and Rechnitzer[8]. The value reported by Ma et al. was considered to be too high, probably due to chains analyzed being too short. Methods used to determine the CAP varied in the literature. Meirovitch and Livne[6] obtained the CAP for SAW in simple cubic lattice with Monte Carlo simulations with the scanning method. They plotted $E(T)/N$ against $N$ and found the exponent $\alpha$ in $E(T)/N \sim N^\alpha$ over three different ranges of chain length ($N$ = 20-60, 60-170, and 170-350). Then, the critical point was located by finding the value of the reciprocal temperature Θ that resulted in the exponent α being constant for the three different ranges of chain lengths. Their reported $\Theta_c$, which is equivalent to our $\varepsilon_c$, was 0.291 ± 0.001. Their method for determining $\Theta_c$ was based on the scaling theory developed by EKB.[4] As stated earlier, at CAP, $E(T)/N$ is expected to scale with $N^{\phi-1}$ where $\phi$ is the crossover exponent. The value of this crossover exponent was debated. EKB first showed that $\phi \approx \nu \approx 0.59$, where ν is the Flory's exponent. Several recent reports suggest that $\phi = 0.5$ even for SAW chains, the same as $\phi$ for random-walks.[8,36] In Meriovitch and Levin's study, $\phi$ was left as an adjustable parameter. The reported $\phi$ value in their study was =0.530± 0.007, slightly larger than recent reported values $\phi$=0.5. If we were to take $\phi$=0.5, then their data would suggest a lower $\Theta_c$. Recently Decase et al.[9] explored four different ways to determine the CAP, mostly based on the scaling idea. They found that a slight change of $\varepsilon_c$ lead to large deviations in the resulting $\phi$. Therefore, simultaneous determination of $\varepsilon_c$ and $\phi$ may not give the true location of CAP. Janse van Rensburg and Rechnitzer[8] studied CAP for SAWs in two and three dimensions using a variety of methods, including studying the energy ratios of walks of different lengths and the specific heats of the chains. They found that analysis of the specific heat data in three dimensions were fraught with difficulty. The energy ratios of different lengths and the free energy method yielded $\varepsilon_c$ within the error bars. They reported a value for the CAP, $\varepsilon_c$=0.288 ± 0.020 and a crossover exponent $\phi$ = 0.5005 ± 0.0036. Our CAP is within the error bars of their reported value. Interestingly, if they assume that the convergence of the energy ratios of different chain lengths is proportional to $1/\sqrt{N}$, the yielded $\varepsilon_c$ = 0.276 ± 0.029, exactly the same as in our study.

The above discussion suggests that the critical condition determined with our approach is the CAP. Our approach to determine the CAP does not depend on knowledge of $\phi$ and therefore does not suffer from the uncertainty in $\varepsilon_c$ when both $\varepsilon_c$ and $\phi$ need to be determined simultaneously. In the remainder of the paper, we will use this method to determine the CAP of SAWs above a planar heterogeneous surface and SAWs for heteropolymers above a planar homogeneous surface.

*4.2. Homopolymers above Heterogeneous Surfaces with Attractive and Non-Interacting Sites*

Here we consider adsorption of homopolymers above a heterogeneous surface. The first type of heterogeneous surface studied consists of a surface composed of two types of sites. One type of the surface sites, which will be called A sites, did not interact with the polymer chains; that is, $\varepsilon_w^A = 0$. The other type of surface site, the B sites, had an attractive interaction with the polymer chains, $\varepsilon_w^B$. The value of $\varepsilon_w^B$ was varied to locate the CAP. Figure 4 shows a plot of the standard deviations in $\beta\Delta\mu^0$ over all chain lengths for each value of $\varepsilon_w^B$ scanned. The minimum in standard deviations occurs for $\varepsilon_w^B(cc) = 0.49 \pm 0.01$, where the error was based on the energy increment scanned. The same method was used to determine the CAP for surfaces with 10%, 15%, 20%, 25%, and 75% attractive sites. Table I summarizes the CAP of homopolymers over heterogeneous surfaces along with the data over a homogeneous surface. Figure 5 presents the plot of CAP, $\varepsilon_w^B(cc)$, as a function of $f_B$ along with the theoretical prediction according to Eq. (8) with $\varepsilon_w^A = 0$ and $\varepsilon_w^h(cc) = 0.276$.

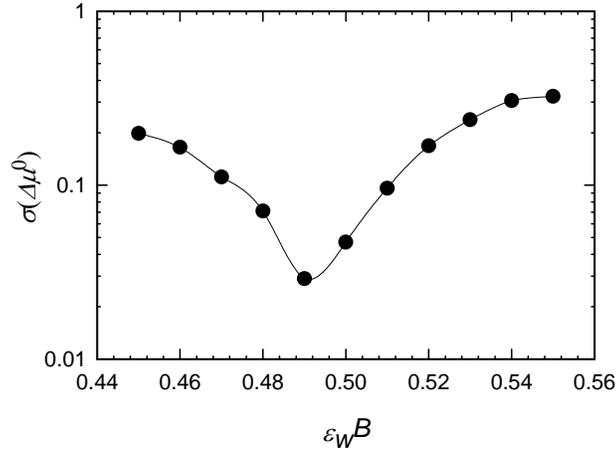

**Figure 4: Plot of deviation in $\Delta\mu^0$ against $\varepsilon_w^B$ for a homogeneous chain adsorbing on a surface with 50% attractive sites and 50% non-interacting sites. The CAP occurs at $\varepsilon_w^B = 0.49 \pm 0.01$.**

It is clear that a good agreement between Eq. (6) and simulation data is observed. Also we note that CAP is not linearly dependent on $f_B$ over the entire range but is well-described by Eq.(6). Earlier study by Sumithra and Baumgaertner[13] focused on surfaces with $f_B$ above the percolation threshold. Within that limited range of $f_B$, a linear dependence may be obtained. This study is the first to confirm the dependence of CAP on the surface disorder over a wide range of $f_B$.

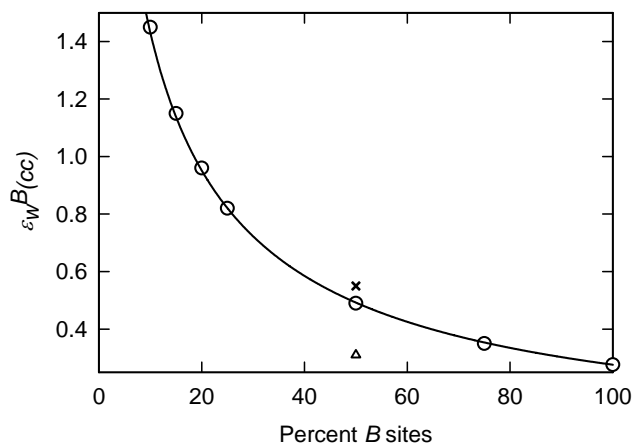

**Figure 5: Plot of the CAP, $\varepsilon_w^B$(cc), against the percent of attractive $B$ sites, $f_B$. The symbols are the CAP determined by the simulation, and the solid line is from equation (6) with $\varepsilon_W^A$ =0.0 and $\varepsilon_w^h$(cc) =0.276. Circles are CAP over random surfaces, the cross (×) is the CAP over a strictly alternating surface, and the upper triangle (Δ) is the CAP over a patchy surface with O.P. =+0.94.**

As discussed in the theory section, one of the assumptions used in deriving Eq. (6) is that the interacting surface sites are randomly distributed. We have tested this assumption by studying adsorption of homopolymers over a 50% surface with alternating and patchy patterns. For a surface with 50% of A and B, an order parameter O.P. can be defined (readers are referred to literature for the definition).[19] If O.P.=0, the surface is random; if O.P.=+1, then the surface is patchy; and if O.P.=-1, the surface is alternating. The data are also included in Table I and are indicated in Figure 4. The two points deviate from the line described by Eq. (6). The CAP obtained over a 50% alternating surface is larger than that over a 50% random surface. On the other hand, the CAP obtained over a 50% patchy surface is smaller than over a 50% random surface. These results can be easily understood. When a chain is adsorbed on the surface, it forms trains, loops and tails.[1] Formation of trains lowers the energy of a chain to overcome the entropy loss during the adsorption. When a chain is in contact with an alternating surface, it is however difficult to form trains as no adsorbing sites are adjacent, while this is possible for random and patchy surfaces. Therefore, chains attraction to the alternating surface is lessened, and adsorption over a 50% alternating surface has to occur at a larger value of $\varepsilon_w$. On the other hand, a chain over a patchy surface can selectively sample patches of the surface composed of adsorbing sites, so the adsorption over patchy surface can occur at a smaller value of $\varepsilon_w$.

Another assumption used in deriving Eq. (6) is the annealed approximation. This approximation is strictly met if the surface pattern in contact with the chain changes during the chain adsorption,[11] hence averaging over different distributions can be performed as done in Eq. (3). The surface in this case is said to contain annealed randomness. If the surface pattern can not change, then the surface is said to contain quenched randomness. In our simulations, the surface contains quenched randomness. In fact, we have used only one realization of a quenched random surface. However, the chain was placed randomly over different surface sites, making the annealed approximation applicable to our simulations. We note that Sumithra and Baumgaertner[13], in their studies, averaged over 50 different realizations of quenched randomness and they compared the results with that of a single surface realization. They did not find major difference between these two approaches, especially if the temperature is high. Moghaddam and Whittington[16] investigated the difference between the quenched average and the annealed average for homopolymer adsorption on heterogeneous surface and random copolymer adsorption on homogeneous surface. Their data show that there was no difference between the two averages in the case of adsorption on random surfaces but there were differences for adsorption of random copolymers especially at low temperature. It has been argued that quenched and annealed averages are equivalent in cases where the quenched surface is large in comparison with the polymer.[11,15] Polotsky *et. al*[18] have also found that the CAP for quenched and annealed surface disorders are the same. In our simulations, the surface is large in comparison with the size of the polymer, and the attachment of the polymer to the surface occurs at many random places on the surface. Therefore, the chain can effectively interact with many different random arrangements of surface sites, and the system approaches the annealed average.

*4.3. Homopolymers above Heterogeneous Surfaces with All Sites Interacting*

In order to assess whether the equation derived for the CAP of random surfaces was valid in more general cases, random surfaces that contained all attractive sites were prepared. For these surfaces, the polymer-surface interaction for the A sites, $\varepsilon_w^A$, was set at a relatively weak attractive strength, 0.10, and the interaction for the B surface sites was varied to find the CAP. Additionally, surfaces with repulsive A sites ($\varepsilon_w^A = -.10$) were also investigated.

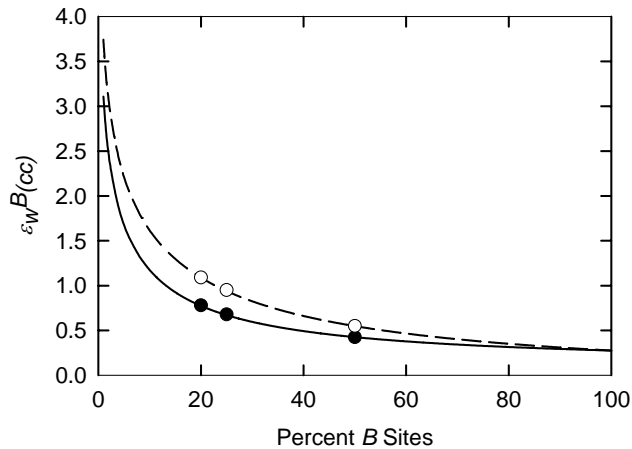

**Figure 6: Plot of the critical adsorption point, $\varepsilon_w^B(cc)$, against the percent of attractive B sites for surfaces with attractive or repulsive A sites. The dashed line and open symbols are for surfaces with slightly repulsive A sites, $\varepsilon_w^A=-0.10$. The solid line and closed symbols are for surfaces with slightly attractive A sites, $\varepsilon_w^A=+0.10$. The symbols are simulation results, while the lines are from equation (6) with the corresponding $\varepsilon_w^A$ values.**

Figure 6 shows the values of $\varepsilon_w^B(cc)$ determined for these two cases, as well as the prediction of the value of $\varepsilon_w^B(cc)$ given the values of $f_B$ and $\varepsilon_w^A$ used in the simulation. As can be seen in the figure, there is a good agreement between the data and the equation, indicating that the equation is valid for surfaces with many different types of surfaces, not just surfaces with attractive and non-interacting sites.

*4.4. Random Copolymers above Homogeneous Surfaces*

Critical adsorption point for random copolymers adsorbing on homogeneous surfaces were also determined. In these systems, polymer chains are considered to be composed of two different types of monomers, A's and B's, interacting with a surface composed of only one type of site. B monomers were attracted to the surface, while A monomers do not interact with the surface, i.e. $\varepsilon_w^A = 0$. Table 2 shows the values of the CAP, $\varepsilon_w^B(cc)$, for various values of $f_B$ along with results obtained for homopolymers, alternating copolymers and block copolymers. Here we have used annealed chemical potentials to determine the CAP. Figure 7 presents the plot of $\varepsilon_w^B(cc)$ as a function of $f_B$ along with the theoretical prediction according to Eq. (6) with $\varepsilon_w^A = 0$ and $\varepsilon_w^h(cc) = 0.276$. The data fit the equation well for situations in which sequences are randomly specified. However, similar to homopolymer adsorption on heterogeneous surfaces, the equation does not apply when the chain sequence is not random. For a diblock copolymer, where the first half of the chain is all A monomers while the second half of the chain is all B monomers, a weaker attraction is required to reach the CAP than for a random 50% copolymer chain. An alternating copolymer requires a slightly stronger attraction to reach the CAP. Again, these results can be explained by considering the tendency of forming trains during adsorption. The diblock copolymer is a homogeneous string of adsorbing B monomers attached to a string of A monomers. The B section of the chain is able to interact with the surface like a homogeneous chain, while the A section does not adsorb and slightly repels the chain from the surface, indicating that the value of $\varepsilon_w^B(cc)$ for a diblock chain should be similar to a homogeneous chain on a homogeneous surface. In fact, $\varepsilon_w^B(cc) = 0.30$ for diblock copolymers, a value only slightly higher than for homopolymer adsorption, and much lower than $\varepsilon_w^B(cc)$ for a 50% random copolymer chain. For an alternating chain, consecutive attractive interactions are not possible, resulting in the necessity of a stronger $\varepsilon_w^B(cc)$ than for a random chain.

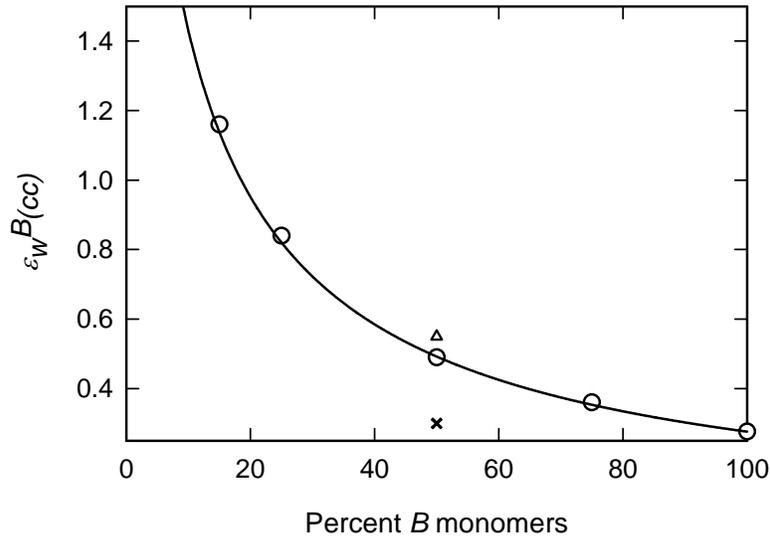

**Figure 7: Plot of the CAP, $\varepsilon_w^B(cc)$, of copolymers over a homogenous surface against the percent of attractive $B$ monomers, $f_B$. The symbols are the CAP determined by the simulation, and the solid line is the plot according to equation (6) with $\varepsilon_W^A = 0.0$ and $\varepsilon_w^h(cc) = 0.276$. Circles are CAP of random copolymers, the cross (×) is the CAP of block copolymers, and (Δ) is the CAP of alternating copolymers.**

Finally we compare the chemical potential determined with annealed approximation versus quenched average. We found that the chemical potential of a random copolymer above the surface, $\mu^0_{ads}$, obtained via the annealed average in Eq. (10) was smaller than the quenched average in Eq. (11). This has been suggested in the literature.[23] Annealed approximation implies that the chain sequence can change when it interacts with the surface. As a result, the chemical potential is lowered when compared with a chain with a fixed sequence. Figure 8 below shows the distribution of $\mu^0_{ads}$, obtained based on trial insertions of a given random sequence, against the sequence order parameter $\lambda$. As discussed in section 3, a generated random sequence may not correspond to exactly $\lambda=0$, therefore resulting a distribution of $\mu^0_{ads}$ against $\lambda$. Figure 8 shows that within the range of $\lambda$ spanned by random sequences, the chemical potential is seen to depend on $\lambda$. The $\mu^0_{ads}$ is higher for negative $\lambda$ and is lower for positive $\lambda$. This is consistent with the results in Table II. A negative $\lambda$ implies the random copolymer chain exhibits statistically alternating behaviour. A higher $\mu^0_{ads}$ implies that the chain is more difficult to be adsorbed on the surface; therefore, it needs a stronger attraction to reach CAP.

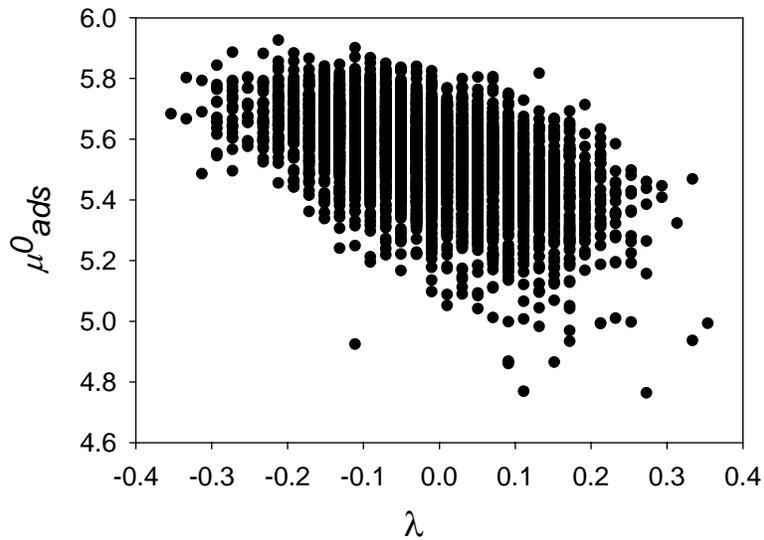

**Figure 8: The distribution of $\mu^0_{ads}$ versus the sequence order parameter $\lambda$ of a random copolymer. Each data point represent one $\mu^0_{ads}$ based on insertion of one given random sequence for 5000 times and the figure contains data for 5000 random sequences. Chain length $N=100$, $f_A = f_B = 0.5$, and $\varepsilon_w^A = 0.0$ and $\varepsilon_w^B = -0.5$.**

## 5. Summary Remarks

Polymer adsorption at surfaces is relevant to many practical applications and has thus received extensive experimental investigation. However, interest in the CAP, to a large degree, has, until recently, remained a theoretical exercise. There were neither experimental methods that directly measure the CAP, nor were there applications that depended on the exact location of the CAP. This has now changed as interesting applications in liquid chromatography separations have been developed.[37,38] In particular, liquid chromatography at the critical condition (LCCC), first reported in the 1980's, has now widely used for characterization of polymer systems that contain structural and chemical heterogeneities. The critical condition in LCCC experiments was defined as the point where homopolymers of a specific type co-elute regardless of their molecular weights. By erasing the dependence of elution on the molecular weights of one species, other species, differing either chemically or structurally, can then be analyzed. Experimentalists[39] have mostly regarded this critical condition as the CAP. Our earlier Monte Carlo simulations largely support this view.[31-34] The current study provides knowledge on the dependence of CAP on sequence disorder or surface disorder and such knowledge will be useful to develop chromatographic methods for analyzing random copolymers.

We note that several earlier studies [13,14,16,17] have examined the adsorption of polymers on surfaces with either surface disorder or sequence disorder. These studies examined influence of disorder on a variety of properties related to polymer adsorption, such as the change of heat capacity, energy of the chain, and radius of gyration of the chain. Very few, however, have tried to determine the dependence of CAP on the disorder. One of possible reasons that hamper these earlier studies to study the dependence of CAP on the disorder may be due to the lack of a convenient way to determine the CAP. As we have discussed in the theory section, CAP was typically understood as the phase transition of an infinitely long chain near a surface. Earlier studies trying to determine the CAP need to wrestle with the difficulty in extrapolation of results to the limit of infinitely long chain. On the other hand, validity of our studies hinges on the way we determine the CAP. In the case of adsorption of homopolymers over homogeneous surface, we discussed the relationship between the CAP determined by our method with reported literature values. Abundant evidence that supports the validity of our approach was presented in section 4.1. However, for the adsorption over heterogeneous surface, the nature of this CAP is not well-understood. Can a long chain in contact with a surface with few adsorbing sites still exhibit a phase transition similar as that of homopolymers over homogeneous surface? If it does, is the transition first-order or second-order? These questions therefore may cast some doubt on the CAP determined by our approach in the presence of disorder. However, the CAP we determined is directly related to the critical condition point in LCCC. Hence, even though the physical meaning of the CAP determined in this study in the presence of disorder could be subjected to further scrutiny, the importance of our results is not undermined.

**Table 1: Critical Adsorption Point for Homopolymers above Heterogeneous Surfaces with Attractive B Sites and Non-interacting A Sites.**

| Percentage of Attractive Sites | $\varepsilon_w^B(cc)$ |
|---|---|
| 100% | $0.276 \pm 0.005$ |
| 75% | $0.35 \pm 0.01$ |
| 50% | $0.49 \pm 0.01$ |
| 25% | $0.82 \pm 0.01$ |
| 20% | $0.96 \pm 0.01$ |
| 15% | $1.15 \pm 0.01$ |
| 10% | $1.45 \pm 0.01$ |
| 50% alternating surface | $0.55 \pm 0.01$ |
| 50% patchy surface (O.P=0.94) | $0.31 \pm 0.01$ |

**Table 2: Critical Adsorption Point for Heteropolymers with Attractive B Monomers and Non-interacting A Monomers over Homogeneous Surface**

| Percentage of B monomers | $\varepsilon_w^B(cc)$ |
|---|---|
| 100% | $0.276 \pm 0.005$ |
| 75% | $0.36 \pm 0.01$ |
| 50% | $0.49 \pm 0.01$ |
| 25% | $0.84 \pm 0.01$ |
| 15% | $1.16 \pm 0.01$ |
| 50% alternating copolymers | $0.55 \pm 0.01$ |
| 50% block copolymers | $0.30 \pm 0.01$ |

Graphics to be used for the Table of Contents

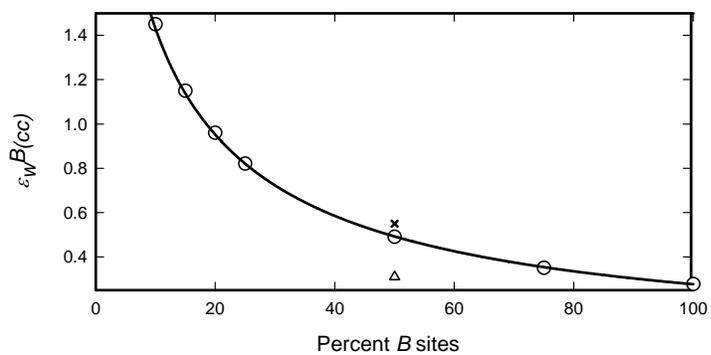